\RequirePackage{lineno}
\documentclass[aps,prc,showpacs,superscriptaddress,reprint]{revtex4-2}
\usepackage{rotating}
\usepackage{epsfig}
\usepackage[colorlinks=true, allcolors=blue]{hyperref}
\usepackage{lineno}
\usepackage{amsmath}
\usepackage{graphicx}
\usepackage[caption=false]{subfig}
\usepackage[dvipsnames]{xcolor}

\usepackage[normalem]{ulem}

\usepackage{xcolor}
\usepackage{xparse,xcoffins}

\ExplSyntaxOn
\NewCoffin\imagecoffin
\NewCoffin\labelcoffin

\keys_define:nn { fqwang/label }
 {
  label   .tl_set:N = \l_fqwang_label_tl,
  labelbox .bool_set:N = \l_fqwang_label_box_bool,
  labelbox .default:n = true,
  fontsize .tl_set:N = \l_fqwang_label_size_tl,
  fontsize .initial:n = \normalsize,
  pos .choice:,
  pos/nwHIGH .code:n = \tl_set:Nn \l_fqwang_label_pos_tl { left,upHIGH },
  pos/nwHigh .code:n = \tl_set:Nn \l_fqwang_label_pos_tl { left,upHigh },
  pos/nwhigh .code:n = \tl_set:Nn \l_fqwang_label_pos_tl { left,uphigh },
  pos/nw .code:n = \tl_set:Nn \l_fqwang_label_pos_tl { left,up },
  pos/nwlow .code:n = \tl_set:Nn \l_fqwang_label_pos_tl { left,uplow },
  pos/nwLow .code:n = \tl_set:Nn \l_fqwang_label_pos_tl { left,upLow },
  pos/nwLOW .code:n = \tl_set:Nn \l_fqwang_label_pos_tl { left,upLOW },
  pos/swHIHI .code:n = \tl_set:Nn \l_fqwang_label_pos_tl { left,downHIHI },
  pos/swHIGH .code:n = \tl_set:Nn \l_fqwang_label_pos_tl { left,downHIGH },
  pos/swHigh .code:n = \tl_set:Nn \l_fqwang_label_pos_tl { left,downHigh },
  pos/swhigh .code:n = \tl_set:Nn \l_fqwang_label_pos_tl { left,downhigh },
  pos/sw .code:n = \tl_set:Nn \l_fqwang_label_pos_tl { left,down },
  pos/swlow .code:n = \tl_set:Nn \l_fqwang_label_pos_tl { left,downlow },
  pos/swLow .code:n = \tl_set:Nn \l_fqwang_label_pos_tl { left,downLow },
  pos/swLOW .code:n = \tl_set:Nn \l_fqwang_label_pos_tl { left,downLOW },
  pos/n .code:n = \tl_set:Nn \l_fqwang_label_pos_tl { hc,up },
  pos/w .code:n = \tl_set:Nn \l_fqwang_label_pos_tl { left,vc },
  pos/s .code:n = \tl_set:Nn \l_fqwang_label_pos_tl { hc,down },
  pos/e .code:n = \tl_set:Nn \l_fqwang_label_pos_tl { right,vc },
  pos .initial:n = nw,
  unknown .code:n   = \clist_put_right:Nx \l_fqwang_label_clist
                       { \l_keys_key_tl = \exp_not:n { #1 } }
 }

\clist_new:N \l_fqwang_label_clist
\box_new:N \l_fqwang_label_box
\box_new:N \l_fqwang_label_image_box

\NewDocumentCommand{\xincludegraphics}{O{}m}
 {
  \group_begin:
  \tl_clear:N \l_fqwang_label_tl
  \clist_clear:N \l_fqwang_label_clist
  \keys_set:nn { fqwang/label } { #1 }
  \tl_if_empty:NTF \l_fqwang_label_tl
   {
    \fqwang_includegraphics:Vn \l_fqwang_label_clist { #2 }
   }
   {
    \SetHorizontalCoffin\imagecoffin
     {
      \fqwang_includegraphics:Vn \l_fqwang_label_clist { #2 }
     }
    \SetHorizontalCoffin\labelcoffin
     {
      \raisebox{\depth}
       {
        \bool_if:NTF \l_fqwang_label_box_bool
         { \fcolorbox{white}{white}{\l_fqwang_label_size_tl\l_fqwang_label_tl} }
         { \l_fqwang_label_size_tl\l_fqwang_label_tl }
       }
     }
    \SetVerticalPole\imagecoffin{left}{36pt+\CoffinWidth\labelcoffin/2}
    \SetVerticalPole\imagecoffin{right}{\Width-36pt-\CoffinWidth\labelcoffin/2}
    \SetHorizontalPole\imagecoffin{upHIGH}{\Height+6pt-\CoffinHeight\labelcoffin/2}
    \SetHorizontalPole\imagecoffin{upHigh}{\Height-0pt-\CoffinHeight\labelcoffin/2}
    \SetHorizontalPole\imagecoffin{uphigh}{\Height-6pt-\CoffinHeight\labelcoffin/2}
    \SetHorizontalPole\imagecoffin{up}{\Height-12pt-\CoffinHeight\labelcoffin/2}
    \SetHorizontalPole\imagecoffin{uplow}{\Height-18pt-\CoffinHeight\labelcoffin/2}
    \SetHorizontalPole\imagecoffin{upLow}{\Height-24pt-\CoffinHeight\labelcoffin/2}
    \SetHorizontalPole\imagecoffin{upLOW}{\Height-30pt-\CoffinHeight\labelcoffin/2}
    \SetHorizontalPole\imagecoffin{downHIHI}{36pt+\CoffinHeight\labelcoffin/2}
    \SetHorizontalPole\imagecoffin{downHIGH}{30pt+\CoffinHeight\labelcoffin/2}
    \SetHorizontalPole\imagecoffin{downHigh}{24pt+\CoffinHeight\labelcoffin/2}
    \SetHorizontalPole\imagecoffin{downhigh}{18pt+\CoffinHeight\labelcoffin/2}
    \SetHorizontalPole\imagecoffin{down}{12pt+\CoffinHeight\labelcoffin/2}
    \SetHorizontalPole\imagecoffin{downlow}{6pt+\CoffinHeight\labelcoffin/2}
    \SetHorizontalPole\imagecoffin{downLow}{0pt+\CoffinHeight\labelcoffin/2}
    \SetHorizontalPole\imagecoffin{downLOW}{-6pt+\CoffinHeight\labelcoffin/2}
    \use:x{\JoinCoffins\imagecoffin[\l_fqwang_label_pos_tl]\labelcoffin[vc,hc]} 
    \TypesetCoffin\imagecoffin
   }
   \group_end:
 }
\NewDocumentCommand{\setlabel}{m}
 {
  \keys_set:nn { fqwang/label } { #1 }
 }
\cs_new_protected:Nn \fqwang_includegraphics:nn
 {
  \includegraphics[#1]{#2}
 }
\cs_generate_variant:Nn \fqwang_includegraphics:nn { V }

\ExplSyntaxOff
\newcommand{\snn}   {\sqrt{s_{_{\rm NN}}}}
\newcommand{\gevc}  {GeV/$c$}

\newcommand{\pt}    {p_T}
\newcommand{\dg}    {\Delta\gamma}
\newcommand{\os}    {{\rm os}}
\newcommand{\sm}    {{\rm ss}}
\newcommand{\cme}   {{\textsc{cme}}}
\newcommand{\bkg}   {{\rm bkg}}
\newcommand{\pair}  {{\rm pair}}
\newcommand{\phib}  {\phi_\bkg}
\newcommand{\phip}  {\phi_\pair}
\newcommand{\dgb}   {\dg_\bkg}
\newcommand{\dgc}   {\dg_\cme}
\newcommand{\ns}    {n_5/s}
\newcommand{\cc}    {c_4/c_2}

\newcommand{\avfd}  {\textsc{avfd}}
\newcommand{\ampt}  {\textsc{ampt}}
\newcommand{\hydj}  {\textsc{hydjet}}
\newcommand{\rqmd}  {\textsc{u}r\textsc{qmd}}
\newcommand{\hij}   {\textsc{hijing}}

\newcommand\mean[1]{\left\langle#1\right\rangle}

\begin{document}

\title{A higher-harmonic observable for the chiral magnetic effect in heavy-ion collisions}

\author{Han-Sheng Li}
\email{li3924@purdue.edu}
\affiliation{Department of Physics and Astronomy, Purdue University, West Lafayette, IN 47907, USA}
\author{Yu-Shan Chang}
\email{chang761@purdue.edu}
\affiliation{Department of Physics and Astronomy, Purdue University, West Lafayette, IN 47907, USA}
\author{Yi Yang}
\email{yiyang429@as.edu.tw}
\affiliation{Institute of Physics, Academia Sinica, Nankang, Taipei 11529, Taiwan}
\author{Fuqiang Wang}
\email{fqwang@purdue.edu}
\affiliation{Department of Physics and Astronomy, Purdue University, West Lafayette, IN 47907, USA}

\begin{abstract}
%\pacs{25.75.-q,25.75.Ld}% PACS
The chiral magnetic effect (CME) is a phenomenon in which electric charge is separated by a strong magnetic field from local domains of chirality imbalance %and parity violation 
in quantum chromodynamics. 
The CME-sensitive azimuthal correlator difference $\Delta\gamma$ between opposite- and same-sign charged hadron pairs is designed to detect charge separation along the magnetic field, on average perpendicular to the reaction plane. 
However, the search for the CME is hindered by large background contributions to $\Delta\gamma$ from particle correlations coupled with elliptic flow.  
In this work, we explore higher-harmonic components in differential $\Delta\gamma(\phi_{\rm pair})$ as a function of the pair azimuthal angle. 
Such components could arise from event-by-event fluctuations of the magnetic fields throughout the collision zone, in both direction and magnitude.
We show by using heavy-ion physics models that the hexadecapole component of $\Delta\gamma(\phi_{\rm pair})$ is sensitive to the CME and insensitive to physics backgrounds. This could offer a unique observable for the CME that is robust against background contributions.
\end{abstract}
\maketitle

%===============================================================================================
\section{Introduction}
One of the unsettled questions in relativistic heavy-ion collisions is the chiral magnetic effect (CME). It is predicted by quantum chromodynamics (QCD) to exist because of vacuum fluctuations of the topological gluon field, yielding a chirality imbalance of (anti-)quarks in local domains because of quark-gluon interactions. Such a chirality imbalance would produce an electric charge separation under a strong magnetic field because of the charge-dependent magnetic moment of the (anti-)quarks~\cite{Kharzeev:2007jp,Fukushima:2008xe}. %Kharzeev:1998kz,Kharzeev:2004ey,
A strong magnetic field is presumably produced in non-central relativistic heavy-ion collisions~\cite{Skokov:2009qp,Deng:2012pc}, on average perpendicular to the reaction plane (RP) spanned by the beam and the impact parameter direction of the collision. The CME charge separation signal is an excess of positively charged particles in one direction along the magnetic field and an excess of negatively charged particles in the opposite direction~\cite{Kharzeev:2007jp}. 

It is convenient to express particle azimuthal distribution in a Fourier series~\cite{Voloshin:2004vk},
\begin{equation}
    dN_\pm/d\phi \propto 1 + 2v_1\cos\phi + 2v_2\cos2\phi + 2a_1^{\pm}\sin\phi + \cdots\,,
    \label{eq:fourier}
\end{equation}
where $\phi$ is the azimuthal angle of particle momentum vector with respect to the RP and the subscript $\pm$ indicates particle's electric charge sign. 
To lighten notation, we adopt the convention that the RP azimuthal angle is zero and simply refer to $\phi$ as particle's azimuthal angle relative to the RP.
The $v_n \equiv \mean{\cos n\phi}$ parameters are called flow harmonics and are taken as charge independent; specifically, $v_2$ is called elliptic flow.
The charge-dependent $a_1^\pm$ parameters characterize the CME signals in positive and negative charged particles, respectively. In absence of finite-number fluctuations, they are opposite in sign and equal in magnitude in each event, $a_1^+=-a_1^-$; from event to event, their signs are random because of random fluctuations of the vacuum topological charge or chirality, rendering a vanishing average signal, $\mean{a_1^+}=\mean{a_1^-}=0$~\cite{Kharzeev:2004ey}. 
Thus, a commonly used observable is a two-particle correlator~\cite{Voloshin:2004vk},
\begin{equation}
    \gamma=\mean{\cos(\phi_\alpha+\phi_\beta)}\,,
\end{equation}
where $\phi_\alpha$ and $\phi_\beta$ are the azimuthal angles of two particles of interest (POI).
Because of the presence of charge-independent backgrounds, such as effects from global momentum conservation, the difference between opposite-sign (OS) and same-sign (SS) correlators is used in experimental searches for the CME~\cite{Voloshin:2004vk},
\begin{equation}
    \dg \equiv \gamma_\os - \gamma_\sm\,.
\end{equation}
The CME signal presented in the $\dg$ observable is $2a_1^2$~\cite{Jiang:2016wve}.
Experimental results indicate significant signals of $\dg$ in heavy-ion collisions at RHIC~\cite{Abelev:2009ac,Abelev:2009ad} and the LHC~\cite{Abelev:2012pa}.

Unlike the parity-odd $a_1$ parameter in the single-particle distribution of Eq.~(\ref{eq:fourier}), the $\dg$ observable is a two-particle correlator and is parity even. As such, $\dg$ is contaminated by charge-dependent physics backgrounds, such as correlations between daughter particles from a resonance decay, or among particles from the same jet or a back-to-back dijet~\cite{Voloshin:2004vk,Wang:2009kd,Liao:2010nv,Schlichting:2010qia}. 
These correlations are often referred to as cluster correlations. 
Coupled with elliptic flow ($v_2$), they contribute to $\dg$ as 
\begin{equation}
    \dgb \propto \mean{\cos(\phi_\alpha+\phi_\beta-2\phib)}\cdot\mean{\cos2\phib}\,,
    \label{eq:bkg}
\end{equation}
where $\phib$ and $\mean{\cos2\phib} \equiv v_{2,\bkg}$ are the azimuthal angle and $v_2$ of the cluster (i.e.~background source).
This charge-dependent background turns out to be significant, possibly accounting entirely for the observed $\dg$~\cite{Schlichting:2010qia}.
Large efforts have since been invested to eliminate or mitigate this background~\cite{Adamczyk:2013kcb,Adamczyk:2014mzf,Khachatryan:2016got,Acharya:2017fau,STAR:2019xzd,Choudhury:2021jwd,STAR:2021mii,STAR:2023gzg,STAR:2023ioo}, including innovative observables~\cite{Xu:2017qfs,Voloshin:2018qsm,STAR:2021pwb,Feng:2021pgf,Feng:2019pxu,Tang:2019pbl,STAR:2020gky,STAR:2023qyt,Li:2024pue}. 
For comprehensive reviews on the CME and experimental search efforts, the reader is referred to Refs.~\cite{Kharzeev:2015zncReviewCME,Kharzeev:2015knaReviewCME,Huang:2015ocaReviewCME,Zhao:2018ixyReviewCME,Zhao:2019hta,Li:2020dwrReview,Kharzeev:2020jxw,Kharzeev:2024zzm,Feng:2025yte,Li:2025yxx}.

%-----------------------------------------------------------------------------------------
\section{Novel idea}
The $\mean{\dgb}$ in Eq.~(\ref{eq:bkg}) is an average quantity integrated over azimuthal angle, which will hereon be enclosed in brackets $\mean{\cdots}$ for clarity. One can also consider a differential quantity $\dgb(\phib)$ which would be simply a $\cos2\phib$ modulation as $\mean{\cos(\phi_\alpha+\phi_\beta-2\phib)}$ is determined by kinematics largely insensitive to $\phib$. It would have little contributions from higher-order harmonics. (However, see a refined discussion on this point in Sect.~\ref{sec:results}.)
Following the same line of thought, one may consider differential $\dg(\phip)$ as a function of the pair azimuthal angle, $\phip$ (i.e.~azimuthal angle of the total transverse momentum of the pair),
\begin{equation}
    \dg(\phip) = \dgc(\phip) + \dgb(\phib)\,,
\end{equation}
where $\phib\equiv\phip$ for a physical background pair (such as a resonance). 
In an ideal situation where the magnetic field is always perpendicular to the RP, $\dgc(\phip)$ would probably have only a single harmonic component, $\cos2\phip$, just like that for the physics background. However, it is well known that the magnetic field (both magnitude and direction) fluctuates from event to event significantly~\cite{Deng:2012pc}, and the magnetic field is not uniform over the collision zone. These effects are suggested by lattice simulations of equilibrium QCD to yield finite CME signals in local regions even if the CME vanishes globally~\cite{Brandt:2024wlw,Brandt:2024fpc,Endrodi:2024cqn}. Moreover, it is possible that the emission direction of CME-driven particles may not perfectly align with the magnetic field  because of dynamical responses of the axial current to medium evolution, as suggested by the Anomalous-Viscous Fluid Dynamics (\avfd) model~\cite{Shi:2017cpu}. It is therefore highly plausible that the CME signal $\dgc(\phip)$ would have contributions from higher-order harmonic components.

Motivated by the preceding idea, we investigate here possible higher harmonics in $\dg(\phip)$ using heavy-ion collision models. 
Our strategy is to fit the $\dg(\phip)$ from model calculations to the following function,  
\begin{equation}
\begin{array}{llll}
    \dg(x) = c_0 &+\ c_2\cos2x    &+\ c_4\cos4x &+\ c_6\cos6x \\
                 &+\ s_1\sin\;\;x &+\ s_3\sin3x &+\ s_5\sin5x \,.
    \end{array}
    \label{eq:fit}
\end{equation}
(We kept the sine terms for reasons that will be explained later.)
We aim to address two key questions: 
\begin{itemize}
    \item[(i)] Are there higher-order harmonic components arising from the CME, e.g., $c_{4,\cme}\neq0$? and
    \item[(ii)] Do background contributions $\dgb(\phib)$ vanish in higher-order harmonics, $c_{n,\bkg}=0\; (n=4,6,...)$? 
\end{itemize}
If the answers are both affirmative, then higher-order harmonics in $\dg(\phip)$ would constitute a unique background-free signature of the CME.

We note that the idea presented here differs from the usual higher and mixed harmonics, such as $\mean{\gamma_{224}}\equiv\mean{\cos2(\phi_\alpha+\phi_\beta-2\psi_4)}$~\cite{Voloshin:2011mx,Voloshin:2012fv}, $\mean{\gamma_{123}}\equiv\mean{\cos(\phi_\alpha+2\phi_\beta-3\psi_3)}$~\cite{CMS:2017pah,ALICE:2020siw}, and $\mean{\gamma_{132}}\equiv\mean{\cos(\phi_\alpha-3\phi_\beta+2\psi_2)}$~\cite{Choudhury:2019ctw}. These are average correlators that include only background contributions, though not directly related to the background in $\mean{\dg}\equiv\mean{\dg_{112}}$. 
Our idea is to examine the $\dg_{112}(\phip)$ differentially as a function of the pair azimuth and to hopefully identify background-free signatures in the higher harmonics of this differential measure.

%===============================================================================================
\section{Brief model descriptions}
We used \avfd~\cite{Shi:2017cpu,Jiang:2016wve,Shi:2019wzi}, A Multi-Phase Parton Transport (\ampt)~\cite{Lin:2001zk,Lin:2004en}, HYDrodynamic plus JETs (\hydj)~\cite{Lokhtin:2012re,Bravina:2013xla}, and the Ultrarelativistic Quantum Molecular Dynamics model (\rqmd)~\cite{Bass:1998ca,Petersen:2008dd} for our study. 
\avfd\ has the capability to simulate CME signals. 
The other heavy-ion models consist of purely physics backgrounds and %, together with \avfd\ without input CME signal, 
are used to address only the background question.  

The \avfd~\cite{Shi:2017cpu,Jiang:2016wve,Shi:2019wzi} model is developed to simulate the CME. 
The initial axial charge density ($n_5$) is dynamically generated to be proportional to the entropy density ($s$), and the strength is set via the proportionality coefficient $\ns$.
Evolution of the chiral fermion currents in the color electromagnetic field is described perturbatively on top of the neutral viscous fluid background of the quark-gluon plasma (QGP).
The QGP medium evolution is simulated event-by-event by a (2+1)-dimensional viscous hydrodynamics, coupled to a hadronic cascade model in the final stage~\cite{Shen:2014vra,Heinz:2015arc}. 
The ``external'' magnetic field is calculated from the spectator protons, and its direction is allowed to fluctuate event-by-event~\cite{Bloczynski:2012en}.
A modest time evolution is assumed for the decreasing magnetic field~\cite{Jiang:2016wve}.
Electric currents are generated along the magnetic fields to be proportional to the chiral chemical potential determined from $n_5$.

The \ampt\ model (version 2.25t4cu2) is a parton transport model~\cite{Lin:2001zk,Lin:2004en}. It consists of a fluctuating initial condition, parton elastic scatterings, quark coalescence for hadronization, and hadronic interactions. 
The model has been extensively tested to reproduce the transverse momentum ($\pt$) spectra and flows of bulk particles. 
We use the string-melting version of \ampt\ and the model parameters are set to their default values. 

The \hydj++ model (version 2.4) is an event generator for heavy-ion collisions by combining  two independent components of soft and hard physics~\cite{Lokhtin:2012re,Bravina:2013xla}. 
The former is determined by thermal equilibrium parameterized by relativistic hydrodynamics, with separate chemical and kinetic freeze-out conditions.
The hard physics component starts from an initial condition by the PYTHIA model~\cite{Sjostrand:2000wi}, with subsequent parton rescatterings and energy loss in the QGP, followed by hadronization.
The model parameters are tuned to reproduce heavy-ion data %at the LHC 
on charged hadron multiplicity, $\pt$-spectra, and flow.

The \rqmd\ model (version 3.4)~\cite{Bass:1998ca,Petersen:2008dd} is a microscopic transport model simulating heavy-ion collisions by propagating hadrons on classical trajectories, featuring color string excitation and fragmentation and hadronic resonance interactions. It simulates the deconfined regime in QCD without explicitly including a QGP. 
\rqmd\ can satisfactorily describe particle production and elliptic flow at lower RHIC energies but does not fully describe experimental data at the higher end of RHIC energy.
% Version 2.4 is used in our simulation with default model settings.
% The weak-decay products are excluded from our analysis.

%===============================================================================================
\section{Results and discussion}\label{sec:results}
The Au+Au collisions at $\snn=200$~GeV were simulated by \avfd\ for the 30-40\% centrality in Ref.~\cite{Choudhury:2021jwd}. We used this same data sample in our study.
Three values $\ns$=0, 0.1, and 0.2 were used for the axial current over entropy density ratio, representing the strength of the input CME signal 0\%, 10\%, and 20\%, respectively. 
A total of 73, 58, and 30 million events were produced for the three $\ns$ values, respectively.
For physics backgrounds, a total of 400, 630, and 193 million minimum-bias Au+Au events at $\snn=200$~GeV were simulated by \ampt, \hydj, and \rqmd, respectively.
Weak-decay particles are kept stable so no weak-decay daughters are included in our analysis.

The impact parameter direction is fixed at $\psi_{\rm RP}=0$ in all the models. We take this as known for the RP direction in our analysis. Centrality is defined by charged hadron ($\pi^\pm$ $K^\pm$, $p$ and $\bar{p}$) multiplicity within $|\eta|<0.5$. The POIs are taken to be within the acceptance of $|\eta|<1$ and $0.2 < p_{T} < 2$~\gevc, typical of midrapidity detectors like STAR~\cite{Abelev:2008ab} and ALICE~\cite{ALICE:2005vhb}.

\begin{figure}[hbt]
%\begin{minipage}{0.49\textwidth}
    \centering
    \includegraphics[width=0.9\linewidth]{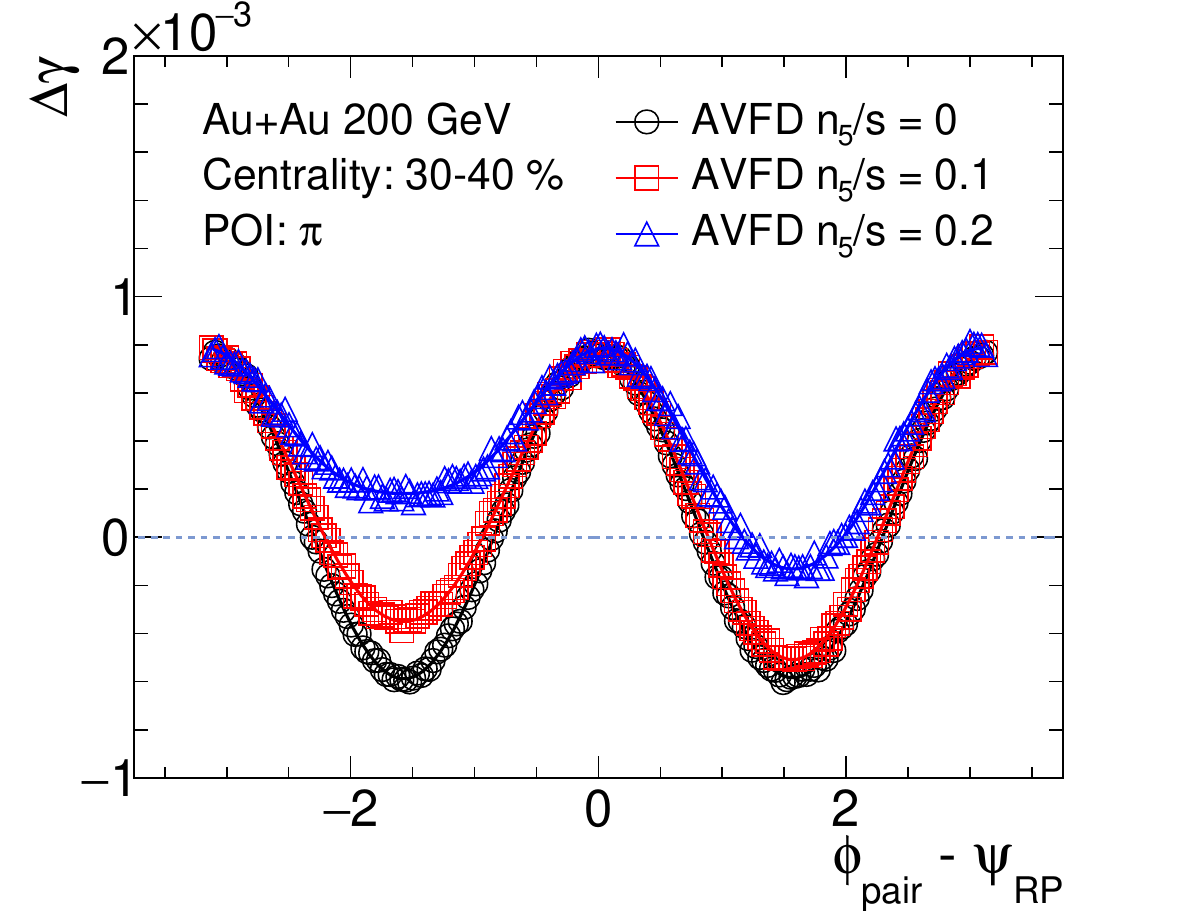}
    \caption{Pion pair $\dg(\phip)$ as a function of pair azimuth in 30-40\% centrality Au+Au collisions at $\snn=200$~GeV simulated by \avfd\ with three cases of axial current, $\ns=0$, 0.1 and 0.2. 
    Superimposed curves are fits to Eq.~(\ref{eq:fit}) with number of degrees of freedom NDF = 193 and $\chi^2=186$, 188, and 204, respectively. We set $\psi_{\rm RP}=0$.
    %Upper panel shows the results with POI: $\pi$, K, p, and lower panel with POI: $\pi$.
    }
    \label{fig:avfd_dg}
%\end{minipage}\hfill
\end{figure}

\begin{figure}[hbt]
%\begin{minipage}{0.49\textwidth}
    \centering
    \includegraphics[width=0.9\linewidth]{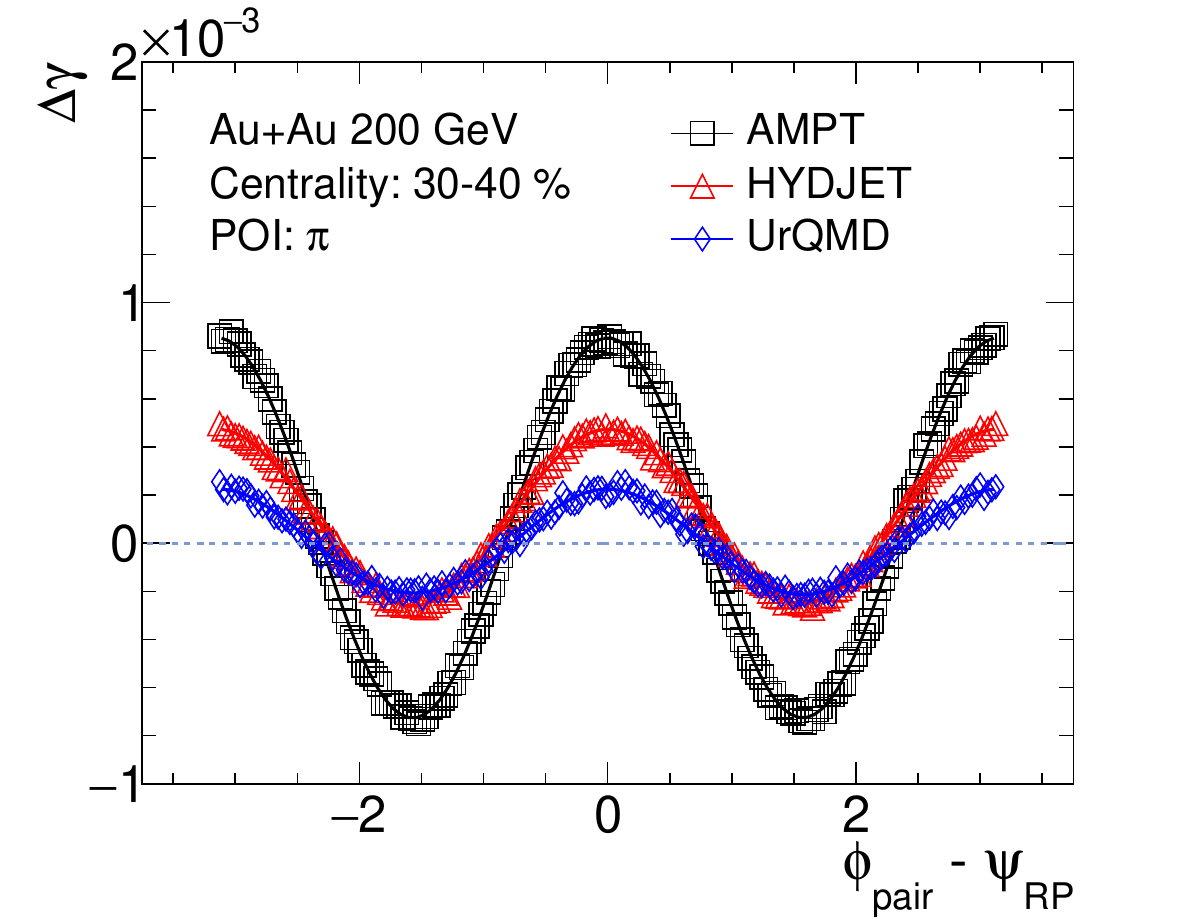}
    \caption{Pion pair $\dg(\phip)$ as a function of pair azimuth in 30-40\% centrality Au+Au collisions at $\snn=200$~GeV simulated by \ampt, \hydj, and \rqmd. 
    Superimposed curves are fits to Eq.~(\ref{eq:fit}) with number of degrees of freedom NDF = 193 and $\chi^2=215$, 273, and 250, respectively. 
    We set $\psi_{\rm RP}=0$.
    }
    \label{fig:bkg_dg}
%\end{minipage}
\end{figure}

Figure~\ref{fig:avfd_dg} shows the $\dg(\phip)$ distributions for the three sets of calculations by \avfd. The $\ns=0$ distribution is symmetric about $\phi_{\rm pair}-\Psi_{\rm RP}=0$. The other two distributions are asymmetric, and this is because \avfd\ artificially fixes the axial current to be single-signed so that $\mean{a_1^\pm}\neq 0$. 
We kept the sine terms in Eq.~(\ref{eq:fit}) to account for these asymmetric distributions.
In reality, the distributions must be symmetric because of the randomness of the topological charge sign.
Figure~\ref{fig:bkg_dg} shows the $\dg(\phip)$ distributions from heavy-ion background models in 30-40\% centrality Au+Au collisions. %The distribution from \avfd\ with no CME input signal is also included in Fig.~\ref{fig:bkg_dg}. 
All distributions are symmetric as expected. 
Figures~\ref{fig:avfd_dg} and \ref{fig:bkg_dg} show results using only pions in computing the correlators; those using charged hadrons are similar.


\begin{figure*}[hbt]
    \centering
    \includegraphics[width=0.333\linewidth]{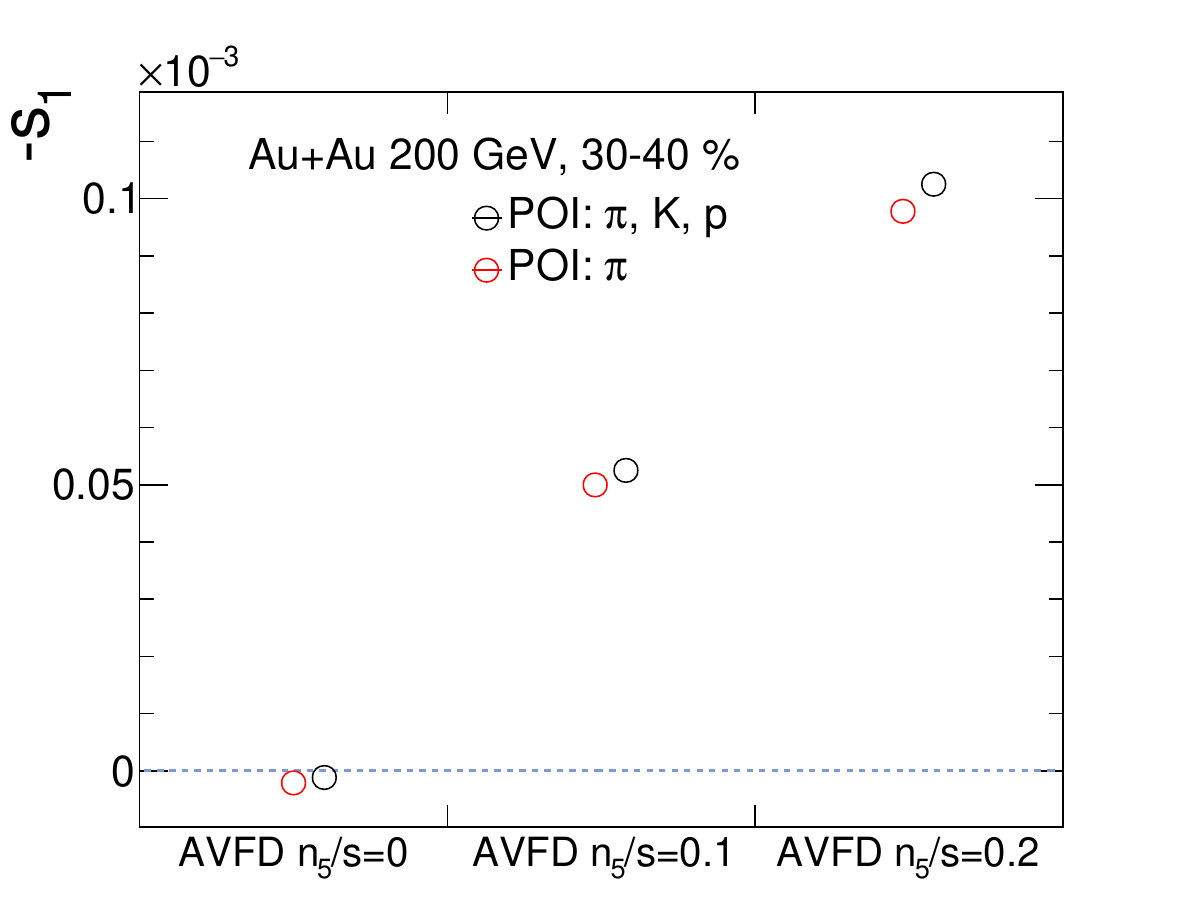}\hfill
    \includegraphics[width=0.333\linewidth]{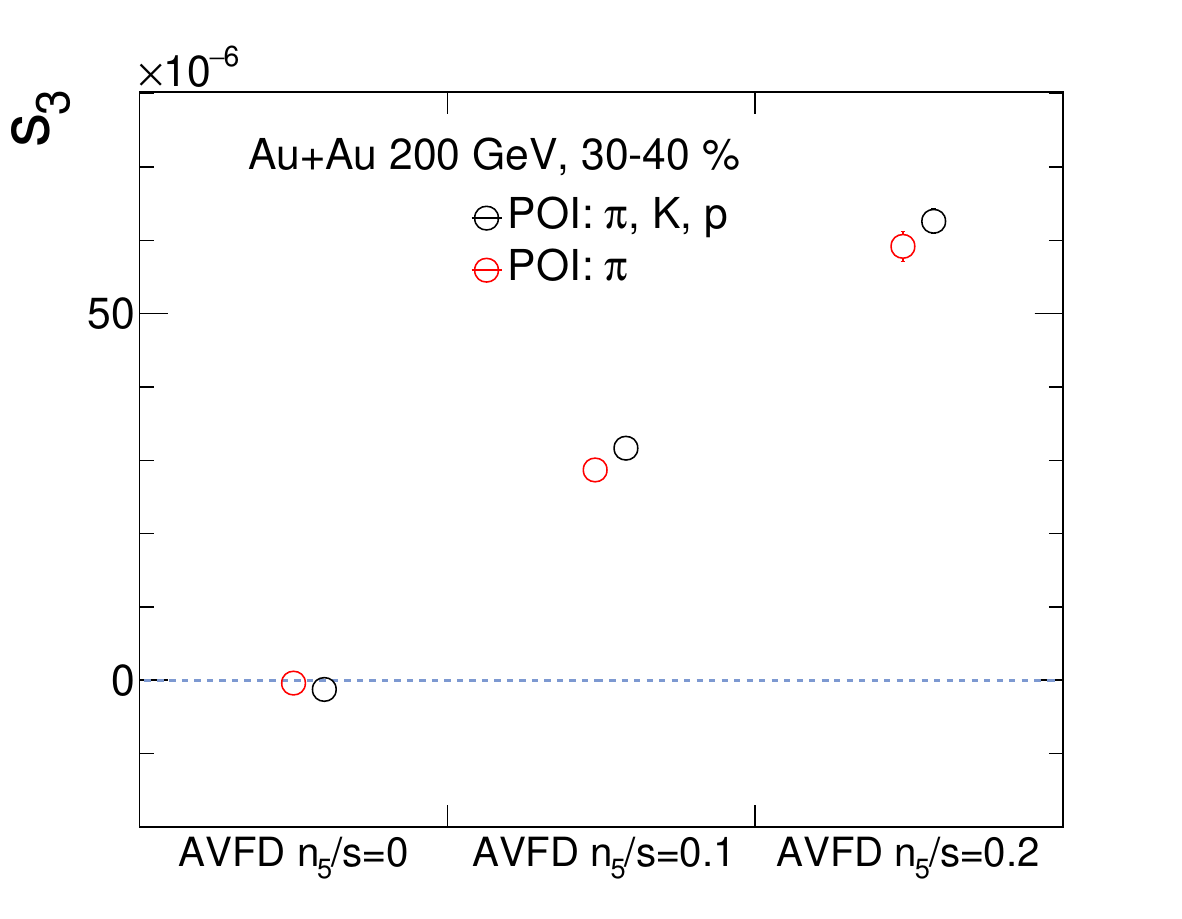}\hfill
    \includegraphics[width=0.333\linewidth]{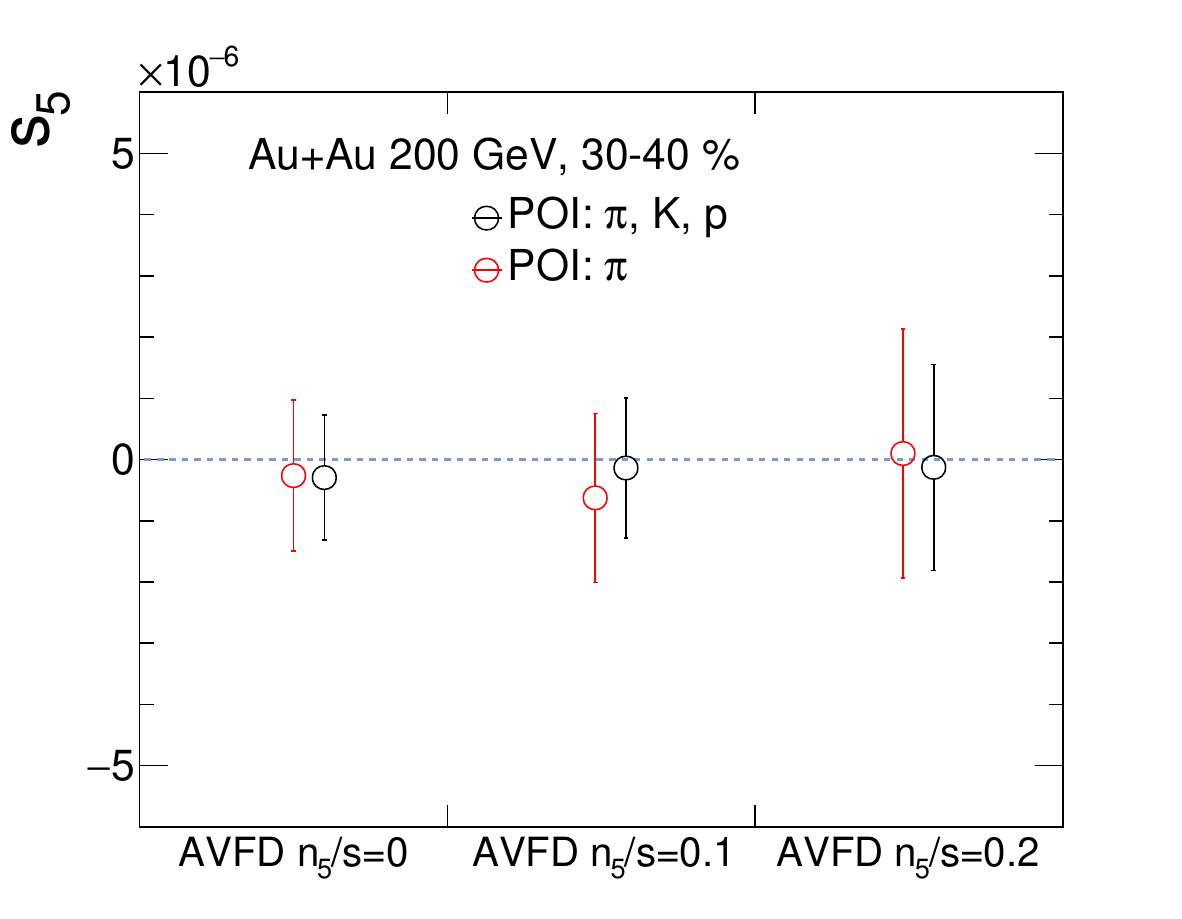}\hfill
    \includegraphics[width=0.333\linewidth]{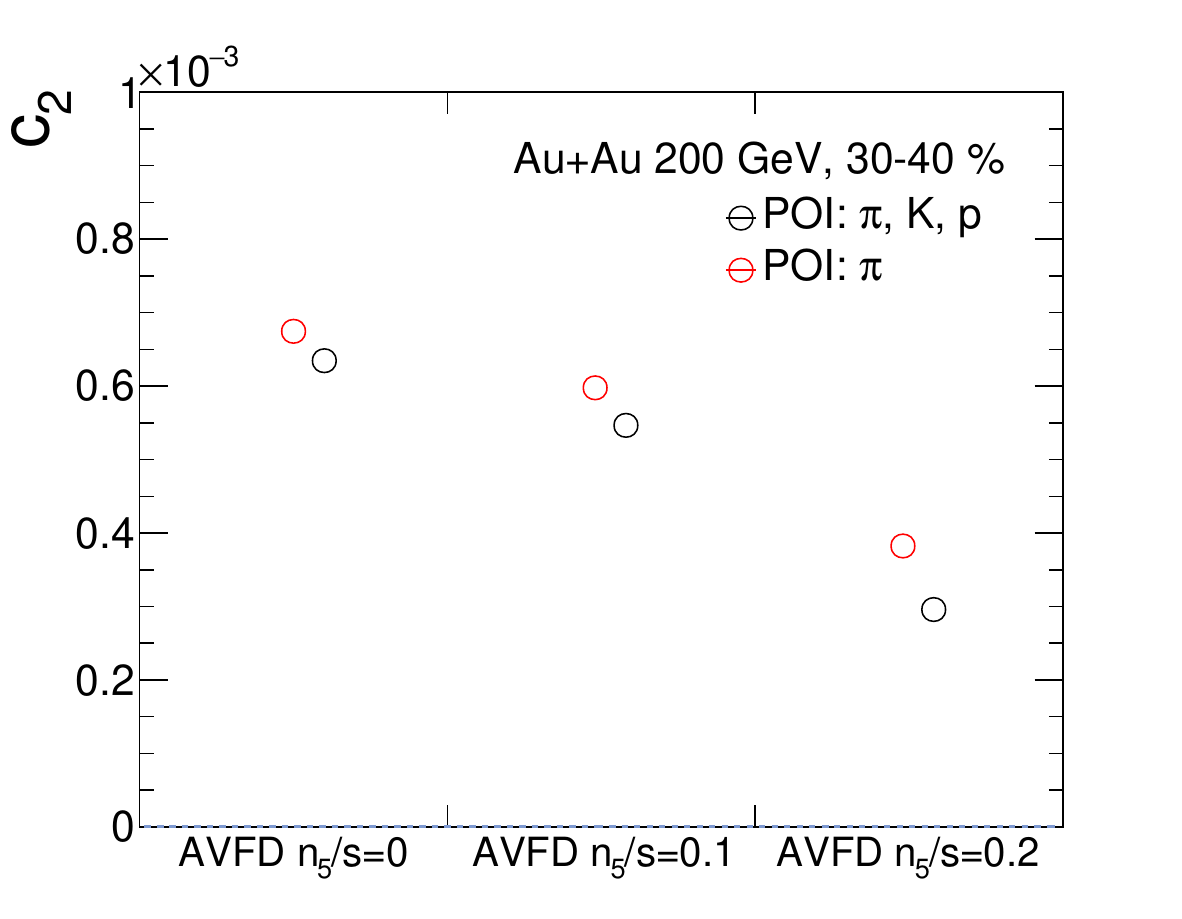}\hfill
    \includegraphics[width=0.333\linewidth]{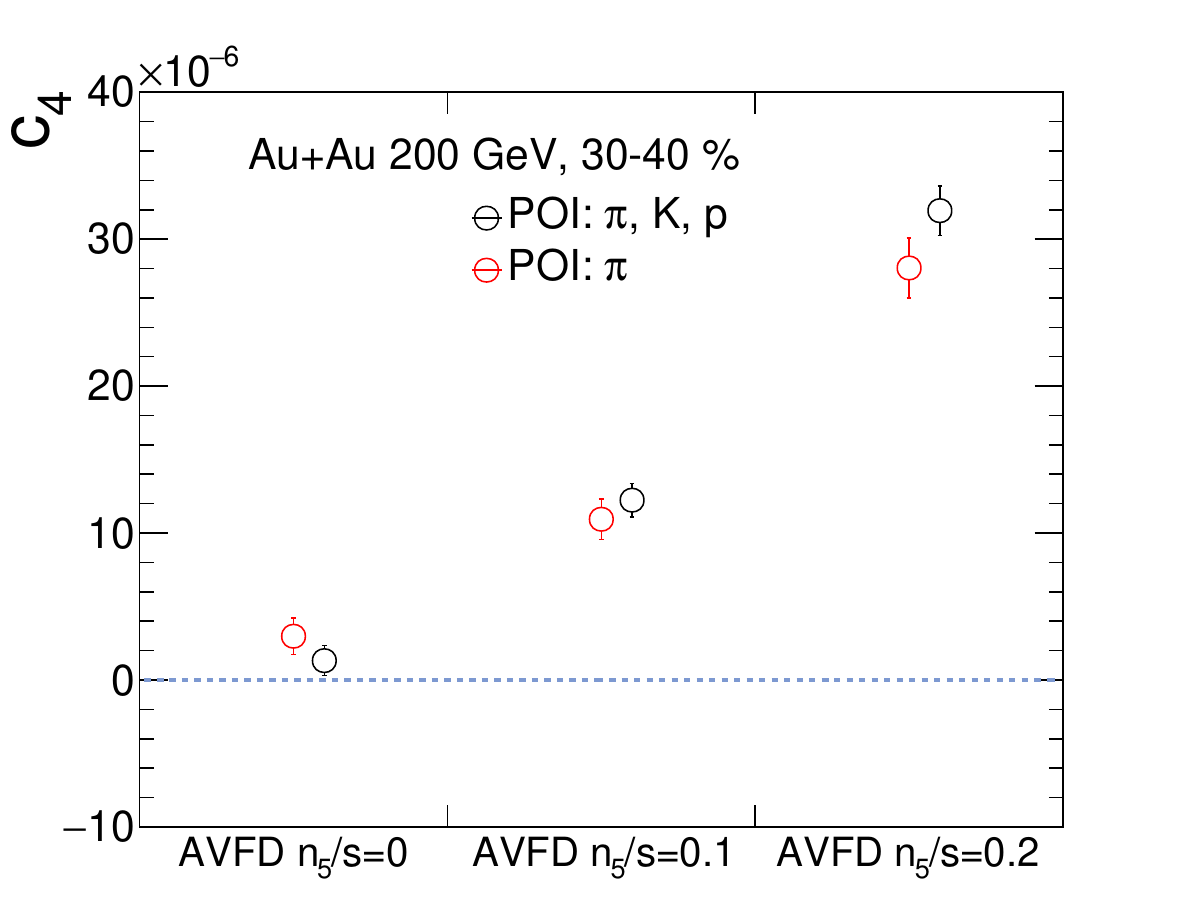}\hfill
    \includegraphics[width=0.333\linewidth]{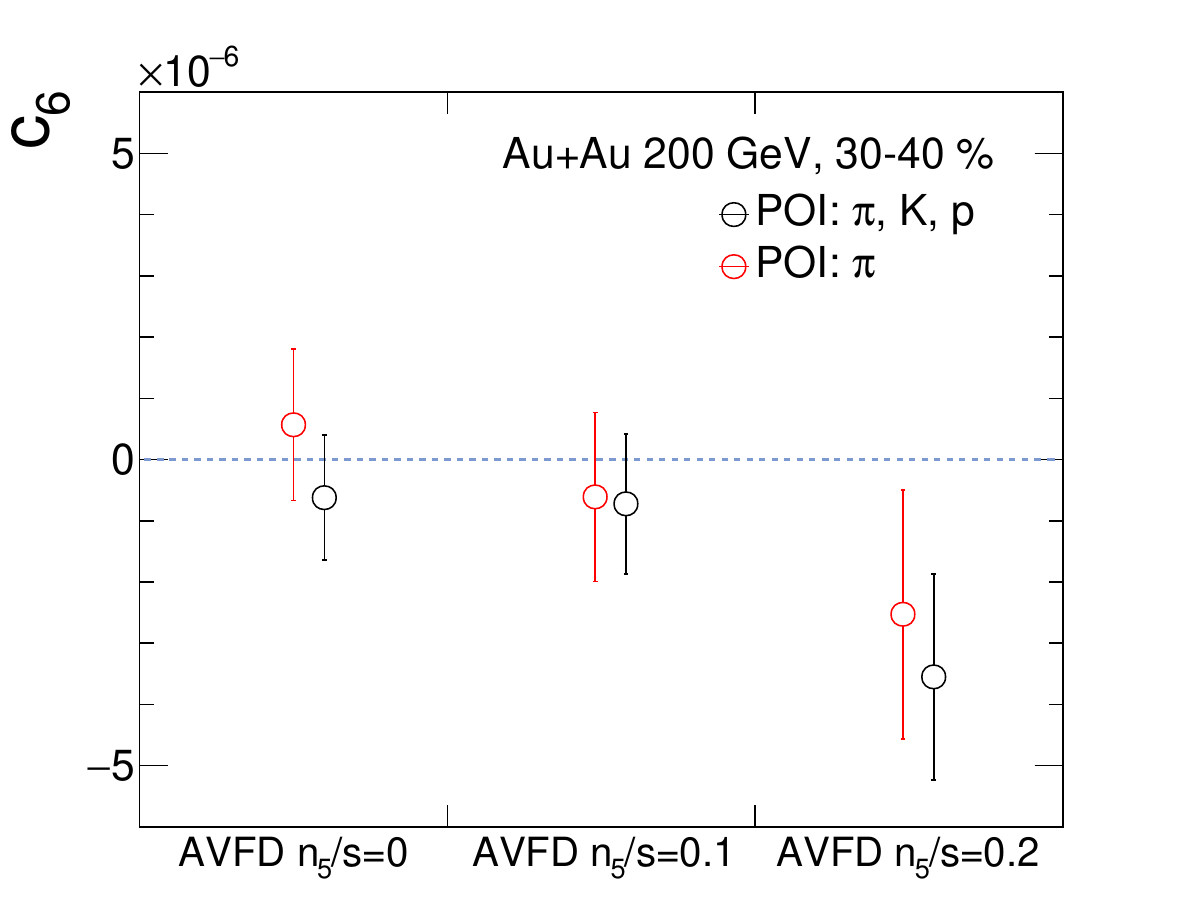}\hfill
    \caption{Fit parameters to \avfd\ simulated data as a function of $\ns$. 
    }
    \label{fig:avfd}
    \centering
    \includegraphics[width=0.333\linewidth]{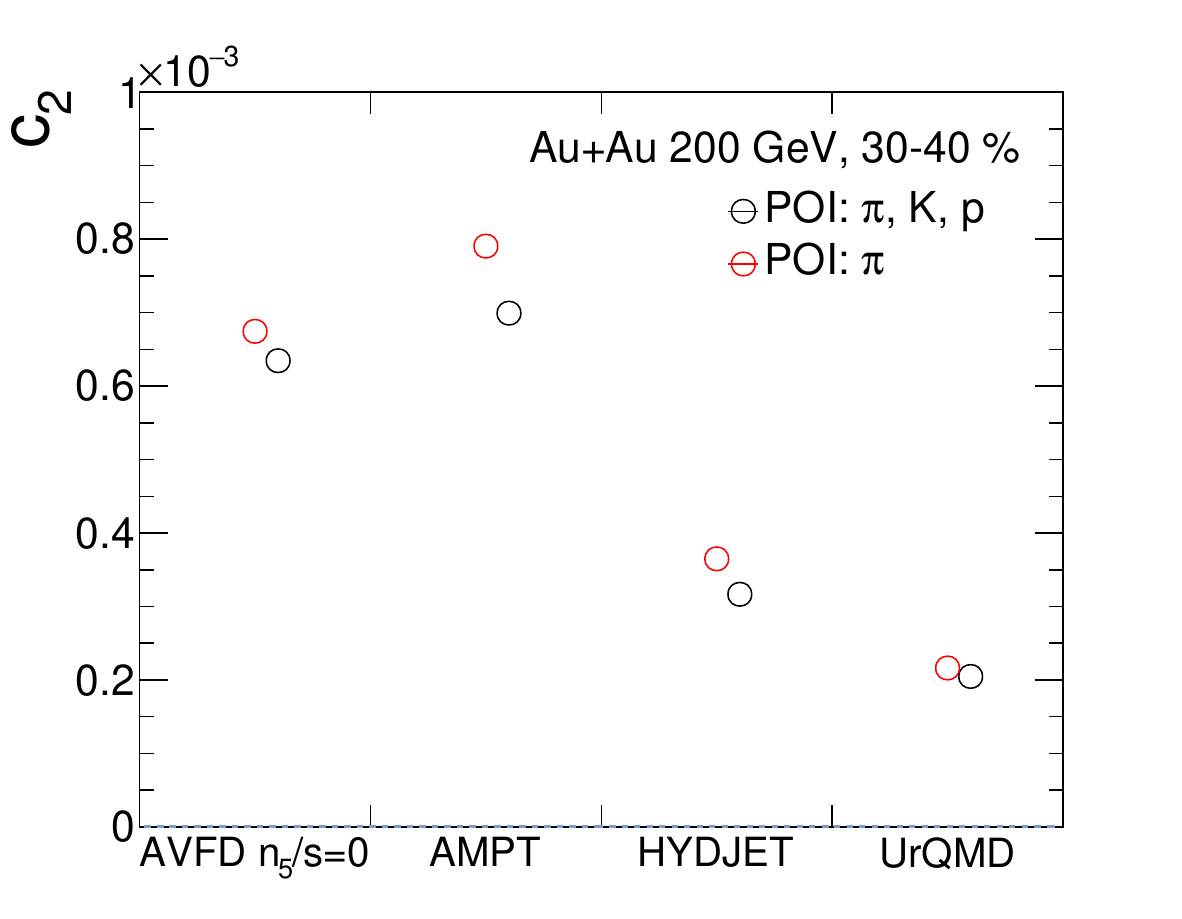}\hfill
    \includegraphics[width=0.333\linewidth]{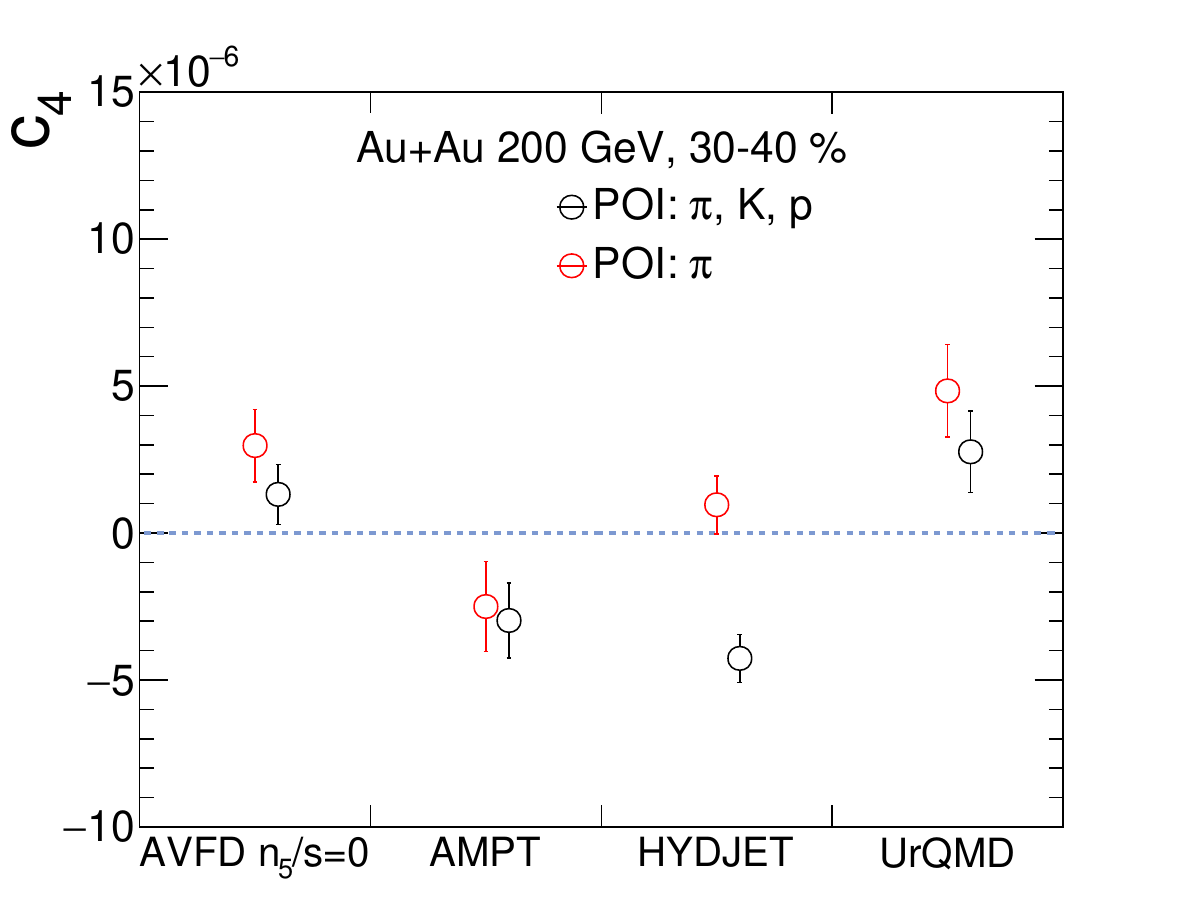}\hfill
    \includegraphics[width=0.333\linewidth]{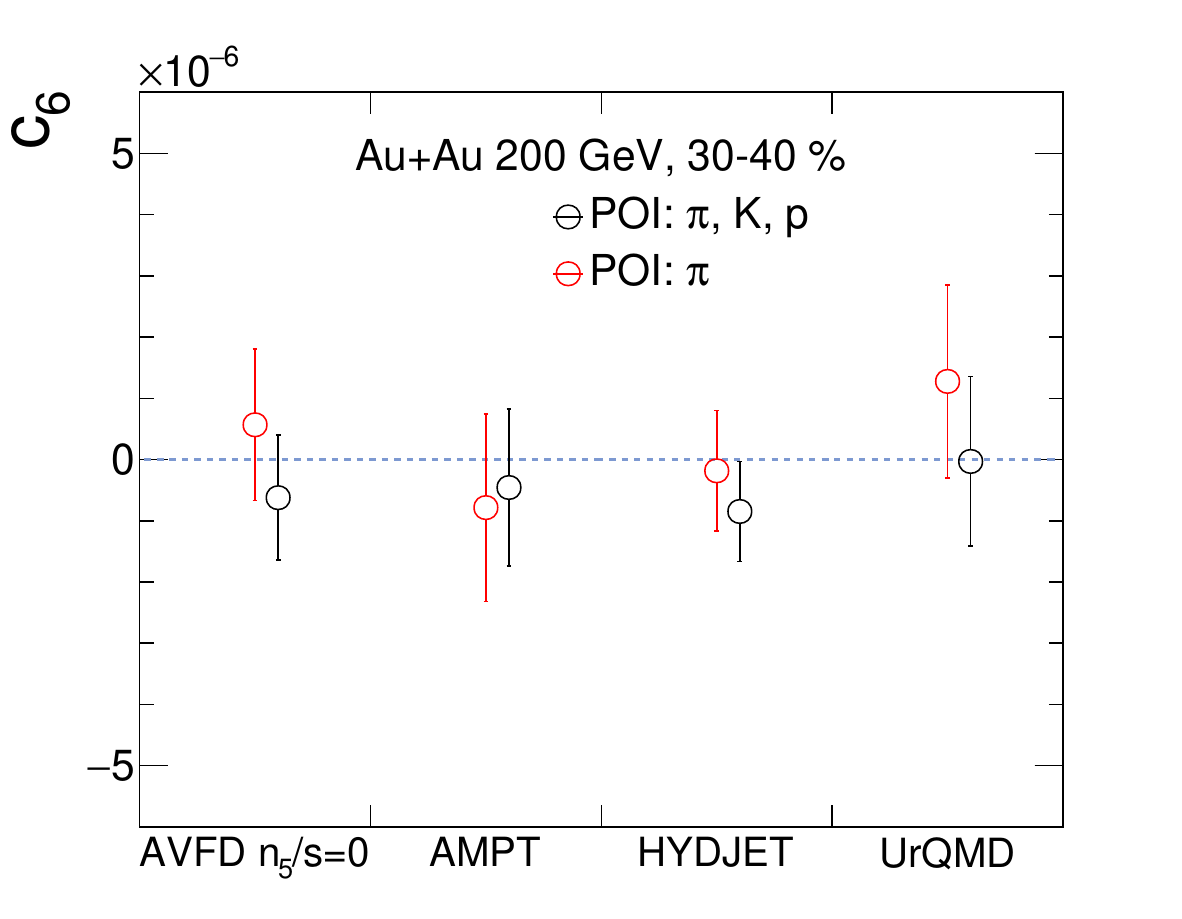}\hfill
    \caption{Fit parameters to simulated data by various physics models, \avfd\ ($\ns=0$), \ampt, \hydj, and \rqmd.
    }
    \label{fig:bkg}
\end{figure*}

The superimposed curves in Figs.~\ref{fig:avfd_dg} and \ref{fig:bkg_dg} are fits to Eq.~(\ref{eq:fit}). The fit parameters are shown in Figs.~\ref{fig:avfd} and \ref{fig:bkg}, respectively, from both using only pions and using charged hadrons ($\pi^\pm$, $K^\pm$, $p$ and $\bar{p}$) which are generally in agreement. 
The $s_5$ and $c_6$ are consistent with zero; we have verified that the harmonics higher than $s_5$ and $c_6$ are all consistent with zero, and this is also reflected in the good quality ($\chi^2/{\rm NDF}\approx 1$) of fits up to harmonics $s_5$ and $c_6$ in Eq.~(\ref{eq:fit}). 

For \avfd, $s_1$ and $s_3$ are nonzero as a result of the artificial $\mean{a_1^\pm}\neq 0$ implemented in \avfd. The $s_1$ and $s_3$ parameters are linear in $\ns$, presumably reflecting the CME signal. Again, in reality, the $s_1$ and $s_3$ components are randomized and the parameters must be zero. For background models, we have verified that all $s_n$ parameters are consistent with zero.
The $c_0$ parameter (not shown) reflects the average $\mean{\dg}$ of the traditional measurement. It increases significantly with increasing $\ns$ in \avfd, reflecting the CME signal strength. The $c_0$ parameter is nonzero from the background models, reflecting background contributions to $\mean{\dg}$.

The $c_2$ parameter is nonzero in both \avfd\ with input CME signals and in background models. These reflect the angular correlation strength of the background sources, $\mean{\cos(\phi_\alpha+\phi_\beta-2\phib)}$, shown in Eq.~(\ref{eq:bkg}), and in the case of \avfd, also contribution from the input CME signal. The $c_2$ parameter is weaker in \hydj\ and \rqmd\ than in \ampt\ and \avfd, presumably because the angular spread of the daughters from the background sources are wider in \hydj\ (near-side minijets) and \rqmd\ (softer resonance production).

\subsection{A novel CME observable: the hexadecapole}
It is intriguing to observe that the hexadecapole $c_4$ parameter is finite in \avfd\ and increases with increasing $\ns$. This indicates that $c_4$ is sensitive to the CME. This finite $c_4$ component is probably caused by the fluctuating magnetic field direction~\cite{Deng:2012pc} and/or the dynamical evolution of the axial current with the medium~\cite{Shi:2017cpu}. Fluctuations in the magnetic field direction will result in the CME being not strictly perpendicular to the RP, likely causing higher-order harmonics in the $\dg(\phip)$ measurement relative to the RP. 
Furthermore, evolution of the CME can depend on the medium density~\cite{Shi:2017cpu} and can be affected by the pathlength-dependent final-state interactions~\cite{Ma:2011uma,Chen:2023jhx}. These effects could result in the maximum CME signal direction not even aligned with the magnetic field, which would also cause higher-order harmonics.

We further observe that the $c_4$ parameter is close to zero in the background models as well as in \avfd\ with no input CME ($\ns=0$). 
The results from using only pions and using charged hadrons agree with other other except those from \hydj. 
We have verified that this is caused by large effects from the baryons ($p$ and $\bar{p}$), presumably because of inadequate descriptions of (anti-)baryon correlations in \hydj. 

It is important to note from Eq.~(\ref{eq:bkg}) that the background modulation should be strictly quadrupolar if the cluster correlations are independent of the cluster azimuth. Such an independency can be broken by azimuthal variations of cluster properties.
For example, resonance $\mean{\pt}$ is modulated because of the $\pt$-dependent elliptic flow, and this can result in the decay correlation $\mean{\cos(\phi_\alpha+\phi_\beta-2\phib)}$ to vary with azimuth.
Such effects are probably responsible for the nonzero $c_4$ in the background models. 

Once cluster correlations are azimuthal dependent, then the factorization used in Eq.~(\ref{eq:bkg}) is no longer valid and the equation in principle does not hold. However, such effects are generally small and Eq.~(\ref{eq:bkg}) still holds approximately. 
Denote $b_\bkg=\mean{\cos(\phi_\alpha+\phi_\beta-2\phib)}$ and suppose $b_\bkg$ is now slightly modulated over azimuth,
\begin{equation}
    b_\bkg = \overline{b}_\bkg(1+2b_2\cos2\phib)\,,
\end{equation}
then a hexadecapole component naturally emerges,
\begin{equation}
    \dg_\bkg(\phib) = \overline{b}_\bkg\cos2\phib + \overline{b}_\bkg b_2\cos4\phib\,,
\end{equation}
where a constant offset is absorbed by $c_0$ in Eq.~(\ref{eq:fit}).
In such a scenario, the ratio $\cc = b_2$ quantifies the quadrupole modulation of the cluster correlation strength.
Such a modulation would be a background contribution to $\cc$.

We thus propose a novel observable, the relative hexadecapole-to-quadrupole $\cc$ strength in the $\dg(\phib)$ distribution, as a signature for the CME. Background contributions to such an observable come from azimuthal dependencies of the angular correlations of background clusters. Such background contributions are of higher order; if they are smaller than the possible CME signal in $\cc$, then our proposed observable can identify the CME.

\begin{figure}[hbt]
    \centering
    \includegraphics[width=0.8\linewidth]{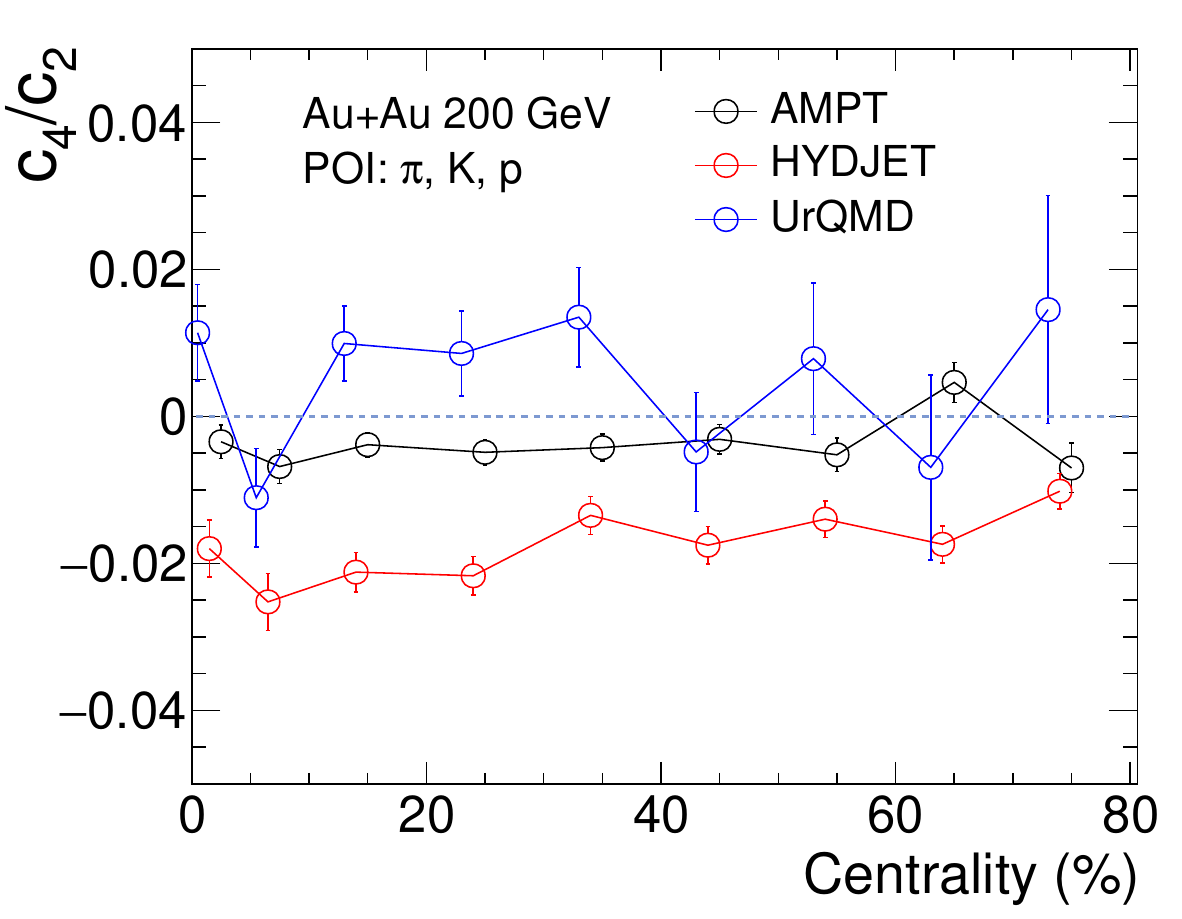}
    \includegraphics[width=0.8\linewidth]{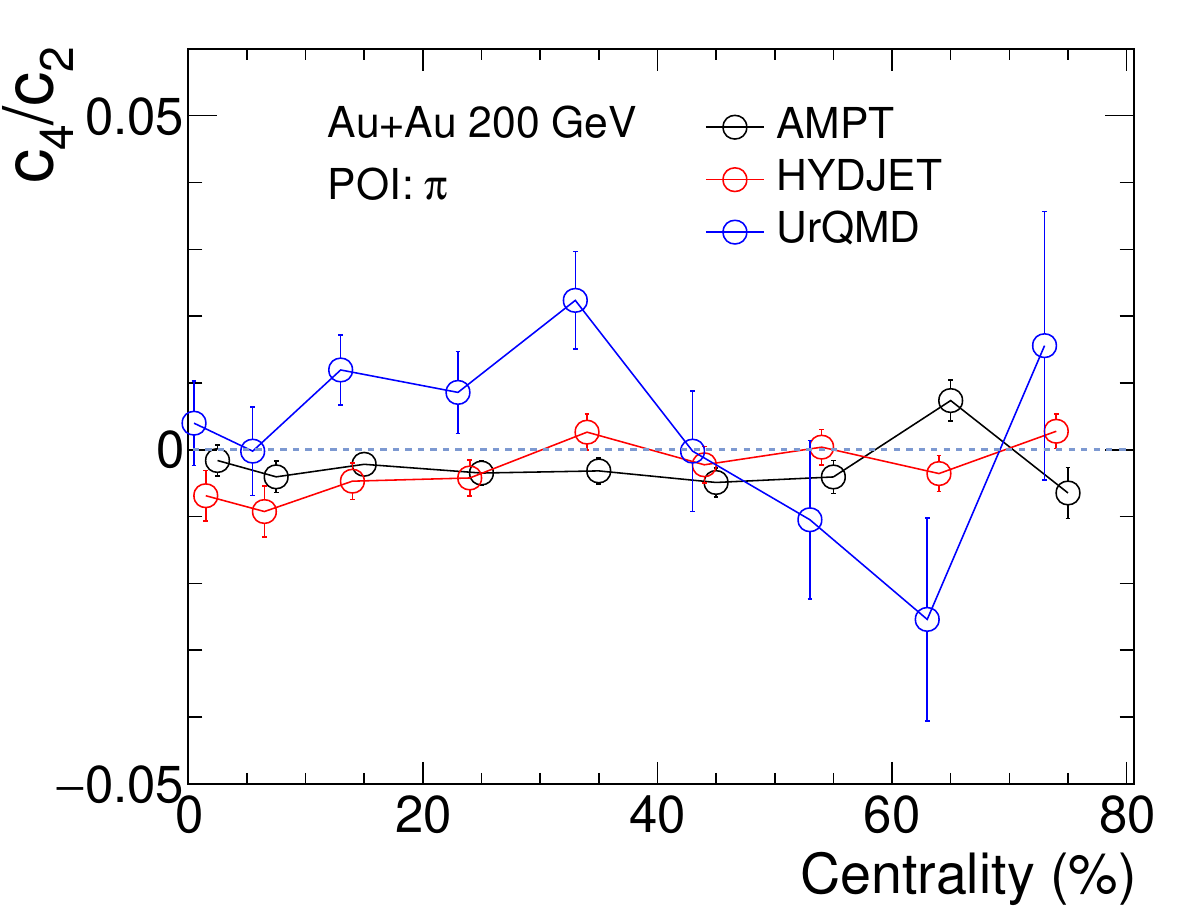}
    \caption{The $\cc$ ratio as a function of centrality in Au+Au collisions simulated by various physics models.}
    \label{fig:c4c2_cent}
\end{figure}

In the following we examine the size of background $\cc$ in the physics models we studied. 
Figure~\ref{fig:c4c2_cent} shows $\cc$ from the background models as a function of centrality for charged hadrons and for pions in the upper and lower panels, respectively. Weak centrality dependence is observed, except for the charged hadron result from \hydj\ previously noted. The background $\cc$ are generally within $\pm 1\%$.%and the $c_4$ parameter is close to zero, mostly within $\pm 10^{-5}$.

\begin{figure}[hbt]
    \centering
    \includegraphics[width=\linewidth,trim={0 1cm 1.5cm 1cm},clip]{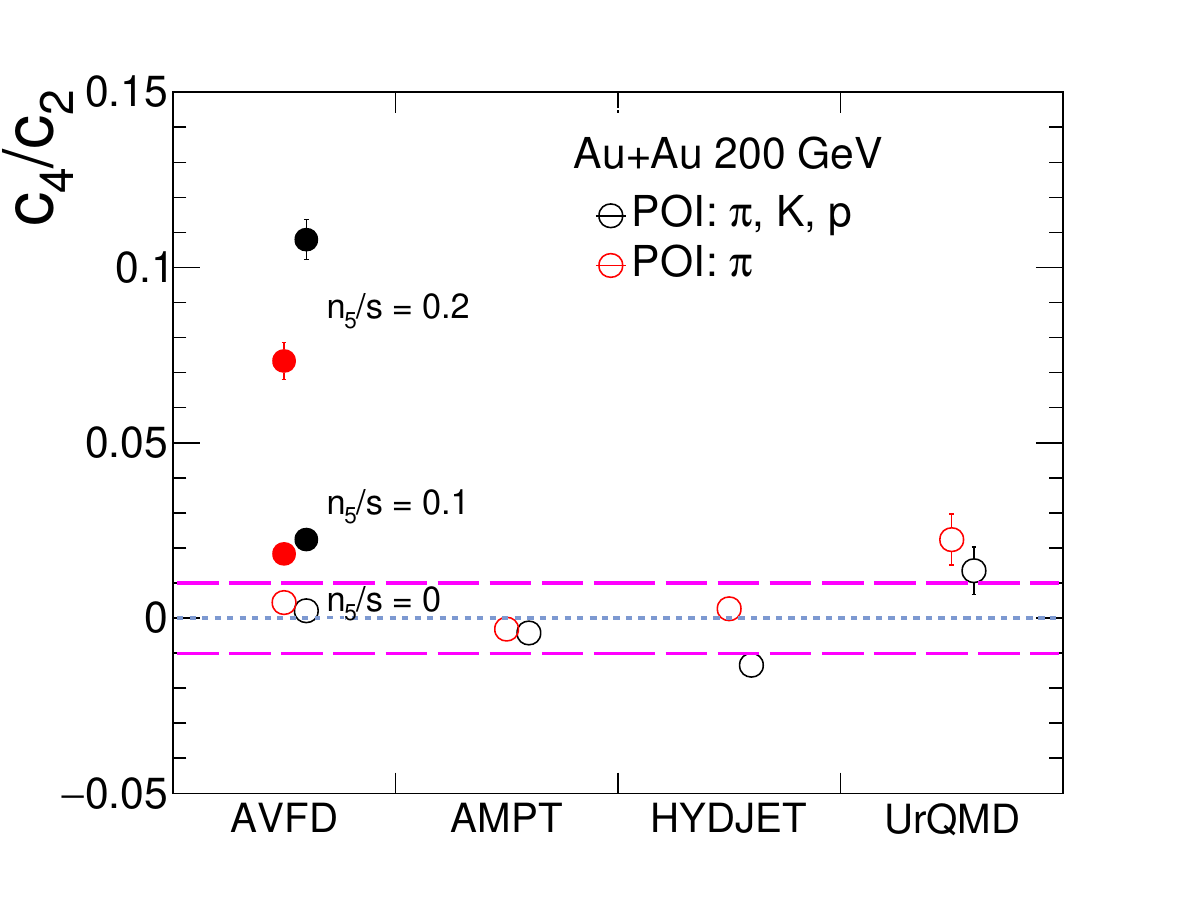}
    \caption{The $\cc$ observable in 30--40\% Au+Au collisions simulated by \avfd\ with and without CME input signals and by physics background models \ampt, \hydj, and \rqmd.}
    \label{fig:c4c2}
\end{figure}

Figure~\ref{fig:c4c2} summarizes the $\cc$ variable from both \avfd\ and background models for the 30--40\% centrality of Au+Au collisions. The region enclosed by the dashed lines at $\pm 1\%$ indicates the background range for charged pions suggested by the studied models. The solid points from \avfd\ with finite CME signals are positive. The $\cc$ parameter appears to increase faster than linearly with increasing $\ns$, similar to the quadratic CME signal $2a_1^2$ as a function of $\ns$~\cite{Shi:2019wzi,Li:2024pue}.
Since the background sources for $\cc$ come in general from azimuthal dependencies of cluster correlations, they should be reasonably well described by available heavy-ion models. 
It thus appears that a CME signal on the order of that corresponding to $\ns=0.1$ in \avfd\ can be identified. 

Because $\dg$ is often measured by the three-particle correlator, $C_3=\mean{\cos(\phi_\alpha+\phi_\beta-2\phi_c)}$, there exists a RP-independent three-particle contribution to $\dg=\Delta C_3/v_2$~\cite{Feng:2021pgf}. A primary source of this contribution is (di-)jet correlations. (Di-)jets can acquire azimuthal modulations from the path-length dependent energy loss, and their correlation shapes can also vary with azimuth because of jet quenching, the net effect of which could contribute to the hexadecapole. We investigated such possible effect using \hij\ (Heavy-Ion Jet INteraction Generator)~\cite{Wang:1991hta,Gyulassy:1994ew}, and found the hexadecapole in $\Delta C_3(\phib)$ is consistent with zero within an statistical uncertainty of $10^{-7}$.
Considering typical $v_2$ values of a few percent, the statistical precision is comparable to $c_4\sim 10^{-6}$, similar to the $c_4$ from \ampt, \hydj, and \rqmd. 
%--------------------------------------------------------------------------------
\section{summary}
Motivated by the general idea that fluctuations tend to produce higher-order effects, we have explored possible manifestations of higher-harmonic components in charge-separation observable in search for the CME, as a result of the strong fluctuations in the magnetic fields generated in relativistic heavy-ion collisions.
Specifically, we investigated the charge-dependent azimuthal correlator $\Delta\gamma(\phi_{\rm pair})$ differentially as a function of the pair azimuthal angle. 
We used the \avfd\ model with the capability of simulating evolutions of chiral fermion currents and the underlying hydrodynamic background to examine higher-harmonic responses to CME signals, and the widely used \ampt, \hydj, and \rqmd\ models to study higher-harmonic responses to physics backgrounds.
It is found that CME signals produce a finite hexadecapole component $c_4$ in the $\Delta\gamma(\phi_{\rm pair})$ distribution, and this component appears to be only weakly sensitive to physics backgrounds.

Background to $c_4$ is generally expected from azimuthal dependent cluster correlations, reflected in the ratio of $\cc$. We thus propose the hexadecapole-to-quadrupole $\cc$ ratio as a viable CME observable. 
The physics background models we studied suggest the background size in $\cc$ to be on the order of $\pm 1\%$.
Since the physics responsible for the background, namely, cluster correlation variation over azimuth, can be reasonably well modeled, the $\cc$ variable provides a viable means to search for the CME. A CME signal on the order of $\ns=0.1$ as implemented in \avfd\ should be identifiable by $\cc$.

\section*{Acknowledgments}
FW thanks Dr.~Gergely Endrődi for valuable discussions that led to the initial idea of this work, 
and Dr.~Jinfeng Liao and Dr.~Shuzhe Shi for discussions on the \avfd\ model. 
This work is supported in part by the U.S.~Department of Energy under Grant No.~DE-SC0012910 (HSL, YSC, FW). 
YY thanks the support from Academia Sinica and the National Science and Technology Council (NSTC) of Taiwan.

\bibliography{refs}% Produces the bibliography via BibTeX.
\end{document}